\documentclass[a4paper,12pt]{article}
\usepackage[dvipsnames]{xcolor}
\usepackage[utf8]{inputenc} 
\usepackage[hidelinks]{hyperref}
\usepackage{float}
\usepackage{amssymb,amsmath,amsthm}
\usepackage{amsfonts}

\pdfoutput=1
\usepackage{graphicx}
\usepackage{graphicx, float, tabularx, booktabs}
\usepackage{color}
\usepackage{enumerate}
\usepackage[american]{babel}
\usepackage{stackrel}

\usepackage[numbers, compress]{natbib}
\usepackage{natbib}
\usepackage{slashed}
\usepackage{amsfonts}
\usepackage{epsfig}
\usepackage{graphicx, tabularx}

\newcommand{\MP}{m_\mathrm{P}}
\newcommand{\LP}{l_\mathrm{P}}

\begin{document}

\begin{center}
{\Large \textbf{How strings can explain regular black holes}}
\end{center}

\vspace{-0.1cm}

\begin{center}
Piero Nicolini

\vspace{.6truecm}

\emph{\small  Dipartimento di Fisica, 
\small Università degli Studi di Trieste, 
Trieste, Italy}\\[1ex]


\emph{\small  Frankfurt Institute for Advanced Studies (FIAS), 
Frankfurt am Main, Germany}\\[1ex]

\emph{\small  Institut f\"ur Theoretische Physik,   
Johann Wolfgang Goethe-Universität Frankfurt am Main, Frankfurt am Main, Germany}\\[1ex]

\end{center}
\begin{abstract}
\noindent{\small  
This paper reviews the role of black holes in the context of fundamental physics. After recalling some basic results stemming from Planckian string calculations,  I present three examples of how stringy effects can improve the curvature singularity of classical black hole geometries. 
}
\end{abstract}

\renewcommand{\thefootnote}{\arabic{footnote}} \setcounter{footnote}{0}
\thispagestyle{empty}
\clearpage


\section{Introduction}
\label{sec:introduction}

Nowadays black holes are the focus of the attention of researchers  working on a variety of topics in Physics and Mathematics.  Astrophysicists have recently observed the shadow of black holes that presumably are harbored in the center of galaxies  \cite{EHT19PR}. Gravitational waves due to black hole mergers have recently been detected at LISA/Virgo facilities \cite{LIV16}. Mathematical physicists and mathematical relativists are interested in the properties of exact black hole solutions \cite{Ren00}. This activity intersects the work of those gravitational physicists  that aim to circumvent the problem of dark sectors by means of theories  alternative to general relativity \cite{SoF10,CaD11,CFP++12}. 
The importance of black hole research, however, goes beyond 
the above research fields. It seems very likely that black holes are fated to be the cornerstone of our understanding of fundamental physics. 

\subsection{Three facts about evaporating black holes}
\label{subsec:threefacts}

To fully appreciate the significance of black holes, it is instructive to go back to the 1970's.
At that time, theoretical physicists were interested in understanding nuclei and their phenomenology. Strings and dual models were formulated just few years earlier and they were already expected to die young due to the advent of QCD. Black holes and general relativity were topics of limited interest, 
because they were disconnected from the quantum realm. Astrophysicists, on the other hand, did not take seriously the existence of black holes, despite the growing evidence accumulated after the initial observation during the suborbital flight of the Aerobee rocket in 1964 \cite{BBCF65}. 
Even the curvature singularity was considered just a mathematical problem, whose solution would never lead to physical consequences. 
 
 Hawking, however,   radically  changed this perspective. He actually set new goals for theoretical physics,  by initiating the study of the Universe from a quantum mechanical view point. Along such a line of reasoning, Hawking showed that, in the vicinity of a black hole, quantum field theory is strongly disturbed by gravity. Particles become an ill-defined, coordinate dependent concept \cite{Ful73,Dav75,Unr76}. To an asymptotic observer black holes appear like black bodies emitting particles at a temperature $T\propto 1/M$, i.e. inversely proportional to their mass \cite{Haw75}.

The existence of a  thermal radiation offered the physical support for the thermodynamic interpretation of the laws governing black holes mechanics \cite{Bek73,BCH73}. It, however, left behind many open questions, such as the fate of an evaporating black hole\footnote{By black hole evaporation one indicate the process of particle emission during the full life cycle. The $1/M$ dependence  implies an increased emission rate as the hole loses mass. Such a nasty behavior is connected to the negative heat capacity of the black hole $C\equiv dM/dT<0$.} and the information loss paradox.\footnote{Microstates of a collapsing star are hidden behind the event horizon. The information is not lost but virtually not accessible. If the hole thermally radiates, it emits particles in a democratic way, de facto destroying the informational content of the initial star. This is the reason why the Hawking radiation is considered an effect that worsens the  problem of the information in the presence of an event horizon. } 
I list below some additional  issues that are too often downplayed:
 \begin{enumerate}[i)]
\setlength{\itemsep}{-\parsep}%
\item If an horizon forms, Minkowski space cannot result from the Schwarzschild metric in the limit $M\to 0$, since it is forbidden by thermodynamics \cite{Wal84};
\item Quantum back reaction effects can tame a runaway temperature \cite{BaB88}, but they can lead to  mass inflation effects
\cite{BaP93};
\label{list:due}
\item Quantum stress tensors imply violation of energy conditions \cite{BiD84}.
\label{list:tre}
\end{enumerate}
In general, issues of this kind  are mostly attributable to a breakdown of Hawking's semiclassical formalism. 
The last item of the above list is, however, intriguing. 
Without energy condition violation,  standard matter would inevitably collapse into a curvature singularity \cite{Pen65}.
As a result, already in the mid 1960's there were proposals, e.g. by Gliner \cite{Gli66} and Sakharov \cite{Sak66}, to improve black hole spacetimes with energy violating source terms. Such proposals culminated with the work of Bardeen, who obtained the first regular black hole solution \cite{Bar68}. The related line elements reads:
\begin{eqnarray}
ds^{2}=-\left(\,1-\frac{2M\LP^{2}\, r^2}{ (r^2
+P^2)^{3/2}} \,\right)dt^{2} +\left( \, 1-\frac{2M\LP^{2}\, r^2}{ (r^2
+P^2)^{3/2}} \, \right)^{-1}dr^{2}+r^{2}d\Omega^{2}.
\label{eq:Bardeenmetric}
\end{eqnarray}
Here the gravitational constant is written in terms of the Planck length $G=\LP^2$. 
 At short scale, the singularity is replaced by a regular quantum vacuum region controlled by a magnetic monopole $P$ \cite{AyG00}.\footnote{Additional regular black hole metrics were proposed in the following years, e.g. \cite{Dym92,AyG99a,AyG99b,AyG99c,Bro01,Mbo05,Hay06,Dym23,Bro23}. For a review see \cite{Ans08}.}  

Against this background, the point \ref{list:tre}) in the above list represented a novelty. The violation was the direct consequence of a major principle, namely the combination of quantum and gravitational effects at short scale. Conversely, for the Bardeen metric, the energy condition violation is the result of an ad hoc choice e.g. the presence of a magnetic monopole. 
For this reason, already in the 1980's semiclassical gravity seemed to pave the way to a possible short scale completion of the spacetime \cite{BiD84}. 

\section{Can one probe length scales smaller than $\sqrt{\alpha^\prime}$?}
\label{sec:canweproble}

As of today, Superstring Theory can be considered the major contender of the ``quantum gravity war'', namely the current debate about the formulation of a consistent quantum theory of gravity. The success of string theory is probably due to its wide spectrum, that covers a vast number of topics and paradigms, from particle physics to cosmology \cite{Nicolai13}. For what concerns black holes, string theory has been applied in a variety of situations, including thermodynamics \cite{Mal99} and derivation of new metrics \cite{Ste98}. The theory has also interesting spin offs where black holes have a major role, e.g.  large extra dimension paradigms \cite{AAD+98,ADD98,ADD99,RaS99a,RaS99b,Gog98a,Gog98b,Gog99,BaF99} and the gauge/gravity duality \cite{Mal99}. There exist also proposals alternative to black holes like the fuzz ball \cite{Mat05}.
String theory is notoriously not free from problems. One of the major limitations is the identification of genuine effective theories, namely the string landscape \cite{Vaf05}. 

\begin{figure}[t]
\includegraphics[width=\textwidth]{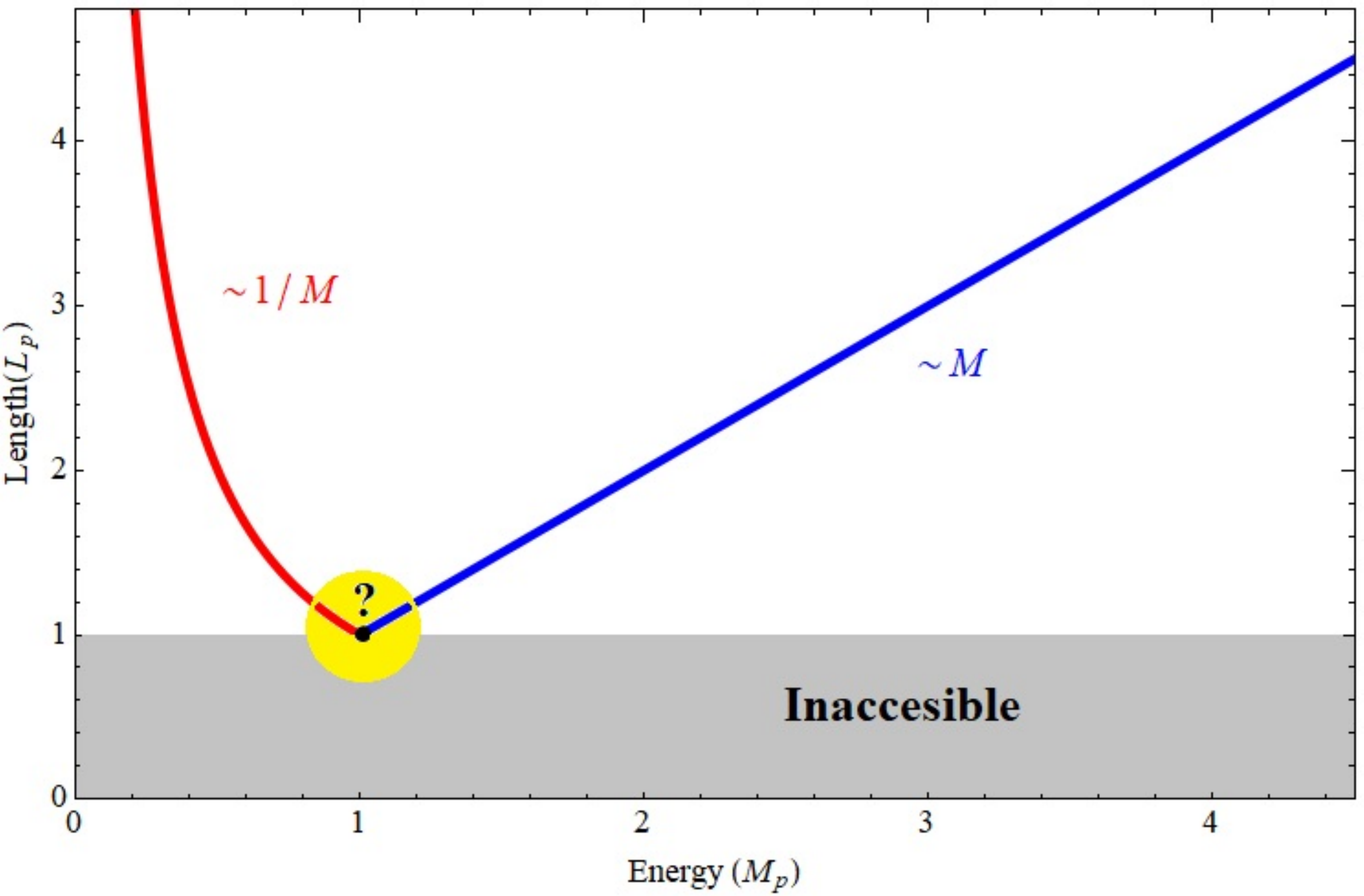}
\caption{The phase diagram of matter at the highest conceivable energy scale.
The red curve represents the Compton wavelength, the blue curve the gravitational radius. The yellow spot represents the regime where string effects are expected to be dominant. The question mark indicates that the nature of the intersection in not known. 
The grey area is virtually inaccessible as a result of the ultraviolet self-complete nature of gravity. 
}
\label{fig:fig1}       
\end{figure}

For the present discussion, it is important to recall just one specif character of string theory: its intrinsic non-locality.
Such a property should come as no surprise, because  strings were introduced to replace quantum field theory and  guarantee ultraviolet finiteness in calculations.  To understand the nature of such a short scale convergence,  string collisions at Planckian energies were extensively studied at the end of the 1980's \cite{ACV87,ACV88,ACV89}. The net result was simple and, at the same time, surprising. The particle Compton wavelength turned to be modified by an additional term, namely:
\begin{equation}
\Delta x \simeq \frac{1}{\Delta p} + \alpha^\prime \Delta p.
\label{eq:GUP}
\end{equation}
Due to the approximations for its derivation,  the above uncertainty relation, known as generalized uncertainty principle (GUP), offers just the leading term of stringy corrections to quantum mechanics. Nevertheless, the GUP can capture several important new features.  
One can start by saying that the GUP depends on the combination of the conventional Compton wavelength $\lambda$ and the gravitational radius $r_\mathrm{g}$ of the particle, being $\alpha^\prime \sim G$. In practice, \eqref{eq:GUP} is a genuine quantum gravity result. The GUP inherits the  non-local character of string theory, being  $\Delta x \geq \sqrt{\alpha^\prime}$. For Planckian string tension $\sim 1/\MP^2$, this is a equivalent to saying that the Planck length $\LP$ is actually the smallest meaningful length scale in nature.  
The GUP also shows that quantum gravity has a peculiar characteristic: Quantum gravity effects shows up \textit{only} in the vicinity of the Planck scale. At energies lower than $\MP$, there is the conventional particle physics. At energies higher than $\MP$, there are conventional (classical) black holes with mass $M\sim\Delta p$. For this reason, one speaks of ``classicalization''  in the trans-Planckian regime \cite{DFG11,DGGK11}. Particles (strings) and black holes are, therefore, two possible phases of matter. The relation between them is evident by the fact that black holes have  a constant ``tension'', $M/r_\mathrm{g}\sim \MP^2$, like a (Planckian) string \cite{AuS13}.   In practice, the GUP suggests that matter compression has to halt due to the gravitational collapse into a Planckian black hole. Such a scenario is often termed gravity ``ultraviolet self-completeness'' and corresponds to the impossibility of probing length scale below $\LP$ in any kind of experiment \cite{Gar95,AuS02,DvG10,Adl10,Car13,Pad20}. The diagram of self-completeness can be seen in Fig. \ref{fig:fig1}.

Despite the great predictive power of the relation \eqref{eq:GUP}, many things remain unclear. For instance, the details of the collapse at the Planck scale is unknown. It is  not clear whether the Lorentz symmetry is actually broken or deformed, prior, during and after the collapse \cite{AuS13}. Also the nature of the confluence of the two curves $\lambda$ and $r_\mathrm{g}$ is debated. One could speculate that there exist a perfect symmetry between $\lambda$ and $r_\mathrm{g}$ and actually particles and black holes coincide \cite{Car13}. Such a proposal, known as ``Black Hole Uncertainty Principle Correspondence'', is currently under investigation and requires additional ingredients for being consistent with observation \cite{CMN23}. Nevertheless, there could be some room for sub-Planckian black holes, as far as there exists a lower bound for the black hole mass \cite{CMN15,CMMN20}. It is also possible to imagine that the confluence is non-analytic \cite{CMMN20}.  
Finally, it has been shown that the number of the dimensions, charge and spin can drastically affect the self-completeness paradigm \cite{CMMN20,MuN13,KKM+19,Nic18}.

\subsection{How to derive a consistent ``particle-black hole'' metric}
\label{subsec:holometric}

There exists at least one thing one knows for sure about gravity self-completeness. The Schwarzschild metric simply does not fit in with the diagram in Fig. \ref{fig:fig1}. The problem is connected to the possibility of having a black hole for any arbitrarily small mass, i.e., $M<\MP$. This implies a potential ambiguity since to a given mass, one could associate  both a particle and a black hole. More importantly, Schwarzschild black holes for sub-Planckian masses have radii $\sim M/\MP^2 $, smaller than $\LP$, a fact that is in contrast with the very essence of self-completeness. The formation of black holes in such a mass regime is the natural consequence of mass loss during  the Hawking emission. Customarily, one circumvents the problem by saying that there is a breakdown of semiclassical gravity. Black holes would explode even before attaining sub-Planckian masses \cite{Haw74}. The problem, however, persists when one considers alternative formation mechanism, like early Universe fluctuations \cite{Haw71,CaH74} and quantum decay \cite{BoH95}.

The most natural way to solve the puzzle is to postulate the existence of an extremal black hole at the confluence of $\lambda$ and $r_\mathrm{g}$. Degenerate horizons are zero temperature asymptotic states that can guarantee the switching off of black hole evaporation.\footnote{The switching off is also known as SCRAM phase, in analogy with the terminology in use for nuclear  power plants \cite{Nic09}.} For instance,  Denardo and Spallucci considered charged black holes and determined the parameters to obtain stable configurations \cite{DeS78}. Microscopic black holes can, however, share their charge and angular momentum very rapidly both via Hawking and Schwinger emissions \cite{Gib75,Pag06}. What one actually needs is a SCRAM phase following the Schwarzschild phase, similarly to what predicted by
Balbinot and Barletta within the semiclassical approximation \cite{BaB88}. 
In conclusion, the issue can be solved only if one is able to derive a metric admitting a Planckian extremal horizon for $M=\MP$. Such a metric exists and it is known as holographic screen metric or simply holographic metric \cite{NiS14}. Its line element reads:
\begin{eqnarray}
ds^{2}=-\left(\,1-\frac{2M\LP^{2}\, r}{ r^2
+\LP^2} \,\right)dt^{2} +\left( \, 1-\frac{2M\LP^{2}\, r}{ r^2+\LP^2  } \, \right)^{-1}dr^{2}+r^{2}d\Omega^{2}.
\label{eq:holometric}
\end{eqnarray}
Eq. \eqref{eq:holometric} is a prototype of a quantum gravity corrected black hole spacetime.  
Indeed, the holographic metric offers a sort of ``preview'' of the characteristics of string corrected black hole metrics, I will present in the next sections. In summary one notice that:
 \begin{enumerate}[a)]
\setlength{\itemsep}{-\parsep}%
\item For $M\gg \MP$, \eqref{eq:holometric} becomes the Schwarzschild metric up to some corrections, that are consistent with the predictions of Dvali and Gomez's quantum $N$-portrait \cite{DvG12,DvG13,DvG13+}; \label{list:one}
\item For $M\simeq 2.06\ \MP$, the Hawking temperature reaches a maximum and the black hole undergoes a phase transition to a positive heat capacity cooling down (SCRAM phase);
\item For $M=\MP$, one has $r_\mathrm{g}=\LP$ and $T=0$, namely the evaporation stops and leaves  a \textit{Planckian extremal black hole} as a remnant;
\item For $M<\MP$,  \eqref{eq:holometric} describes a horizonless spacetime due to  a particle sitting at the origin.
\label{list:four}
\end{enumerate}
In practice \eqref{eq:holometric} perfectly separates the two phases of matter, i.e. particles and black holes,  and protects the region below $\LP$ in Fig. \ref{fig:fig1} under any circumstances.
In addition,  \eqref{eq:holometric} does not suffer from quantum back reaction, being $T/M\ll 1$ during the entire evaporation process. Also the issue of the mass inflation at point \ref{list:due}) in Sec. \ref{subsec:threefacts}  is circumvented. For $M>\MP$, there are actually an event horizon $r_\mathrm{g}=r_+$  and a Cauchy horizon $r_-$,
\begin{equation}
 r_\pm = \LP^2\,\left(\, M \pm \sqrt{M^2 -\MP^2}\,\right),
 \label{eq:holohorizon}
\end{equation}
but the latter falls behind the Planck length and it is actually not accessible. From \eqref{eq:holohorizon}, one notices that the horizon structure is the same of the Reissner-Nordström black hole, provided one substitutes the charge with the Planck mass
\begin{equation}
Q/G\longrightarrow \MP.
\label{eq:selfimplementation}
\end{equation}
Eq. \eqref{eq:selfimplementation} is another key aspect of self-completeness. Gravity does not need the introduction of a cut off. The completeness is achieved by exploiting the coupling constant $G$ as a short scale regulator. At this point, there is, however, a caveat: The spacetime \eqref{eq:holometric} does have a curvature singularity. The regularity was not the goal of the derivation of such a metric. The basic idea has been the introduction  of fundamental surface elements (i.e. holographic screens), as building blocks of the spacetime. Each of such surface elements is a multiple of  the extremal configuration, that becomes the basic information capacity or information bit. Indeed for the holographic metric the celebrated area law reads
\begin{equation}
S(\mathcal{A}_+)=\frac{\pi}{\mathcal{A}_0}\left(\, \mathcal{A}_+-\mathcal{A}_0\,\right)
+\pi\ln\left(\,\mathcal{A}_+/\mathcal{A}_0\, \right)
\label{eq:area-entropy}
\end{equation}
where $\mathcal{A}_0=4\pi\LP^2$ is the area of the extremal event horizon, and $\mathcal{A}_+=n\mathcal{A}_0$. 

If one accepts that surfaces (rather than volumes) are the fundamental objects, the question of the regularity of the gravitational field inside  the minimal surface is no longer meaningful. The spacetime simply ceases to exist inside the minimal holographic screen. This interpretation is reminiscent of spacetime dissolution observed in quantum string condensates within  Eguchi's areal quantization scheme \cite{AAS99a}. \footnote{The classical spacetime is a condensate of quantum strings. At distances approaching  $\sqrt{\alpha^\prime}$, long range correlations of the condensate are progressively destroyed.  For $\alpha^\prime=G$, the whole spacetime boils over and no trace of the string/$p$-brane condensate is left over.}

\section{What T-duality can tell us about black holes}

Suppose one has a physical system living on a compact space, whose radius is $R$. Suppose there exists another physical system defined on another compact space, whose radius is proportional to $1/R$. If the observables of the first system can be identified with that of the second system, one can say that such systems are equivalent or dual with respect to the transformation\footnote{The duality is termed T-duality, or target space duality.}  
\begin{equation}
R\longrightarrow 1/R.
\end{equation}
For example, by setting $R\sim 1/\Delta p$ in \eqref{eq:GUP} one finds
\begin{equation}
\Delta x\simeq R + \frac{\alpha^\prime}{R}.
\end{equation}
The above relation actually maps length scales shorter than $\sqrt{\alpha^\prime}$ to those larger that $\sqrt{\alpha^\prime}$,
being
\begin{equation}
\Delta x(R) = \Delta x (1/R),
\end{equation}
for suitable values of $\alpha^\prime$. From this viewpoint, one can say that the GUP is a T-duality relation. This fact is per se intriguing because it offers an additional argument for a stringy interpretation of the holographic metric. The good part is that T-duality allows for an  even more genuine contact between string theory and a short scale corrected metric.
To do this, one needs to go back to a basic result due to Padmanabhan \cite{Nic22}.

Standard path integrals can be thought as the sum of amplitudes over  all possible particle trajectories. In the presence of gravity, the scenario is slightly modified. Indeed, there exist paths that cannot contribute to the path integral. If paths are shorter than the particle gravitational radius, they must be discarded in the computation of the amplitude.  A simple way to achieve this is to introduce a damping term, 
\begin{equation}
 e^{-\sigma(x,y)/\lambda}\longrightarrow   e^{-\sigma(x,y)/\lambda} e^{-r_\mathrm{g}/\sigma(x,y)},
\end{equation}
for each  path contribution in the sum over the paths.\footnote{We temporarily assume Euclidean signature for the ease of presentation.} The above relation implies that the path length $\sigma(x,y)$ admits a minimum. Interestingly, Padmanabhan performed the  sum over the above path contributions and derived a modified propagator \cite{Pad97,Pad98}
\begin{equation}
G(x,y; m^2)=\int \frac{d^D p}{(2\pi)^D}e^{-ip\cdot(x-y)} G(p),
\label{eq:paddypropcoord}
\end{equation}
with
\begin{equation}
G(p)= -\frac{l_0}{\sqrt{p^2+m^2}}\, K_1 \left(l_0 \sqrt{p^2+m^2}\right),
\label{eq:paddyprop}
\end{equation}
where $K_1(x)$ is a modified Bessel function of the second kind, and $l_0$ is called ``zero point length'' \cite{Pad20}.
Eq. \eqref{eq:paddyprop} is intrinsically non-local: The Bessel function has a damping term for momenta larger than $1/l_0$. Therefore $l_0$ is the minimal length that can be resolved over the manifold. Conversely, for small arguments one finds the conventional quantum field theory result.

The virtue of Padmanabhan's calculation \eqref{eq:paddyprop} is two fold: The propagator  is a robust result that descends from general considerations; The functional form in terms of the Bessel function  exactly coincides with the correction of string theory to standard, ``low energy'' quantum field theory.  To better understand such a crucial point we briefly sketch the line reasoning at the basis of series of papers authored by Spallucci and Padmanabhan in collaboration with Smailagic \cite{SSP03} and Fontanini \cite{SpF05,FSP06}.   

Let us start by considering a closed bosonic string in the presence of just one additional  dimension, that is compactified on circle of length $l_0=2\pi R$. 
The string mass spectrum can be written as
\begin{equation}
M^2=\frac{1}{2\alpha^\prime}\left(n^2\frac{\alpha^\prime}{R^2}+w^2 \frac{R^2}{\alpha^\prime}\right)+\mathrm{harmonic \ excitations},
\end{equation}
where $n$ labels the Kaluza-Klein excitations and $w$ is the winding number of the string around the compact dimension.\footnote{In the process of path integral quantization, harmonic oscillators are irrelevant. Therefore we consider them frozen without unwanted  consequences.} 
As expected the above relation enjoys T-duality. It is invariant under simultaneous exchange $R\leftrightarrow \alpha^\prime /R$ and $n\leftrightarrow w$ and leads to the identification of $\sqrt{\alpha^\prime}$ as invariant length scale.

Strings are intrinsically non perturbative objects. As a result, any perturbative expansion destroys the very essence of the theory. The only way to extrapolate a nonpertubative character that can be ``adapted'' to the field theoretic concept of particle is the study of the string center of mass (SCM) dynamics. From the propagation kernel of the SCM in five dimensions, one can integrate out the fifth dimension to obtain an effective four dimensional propagator
\begin{equation}
K(x-y, 0-nl_0; T)=\sum_n\int [\mathcal{D}z][\mathcal{D}p][\mathcal{D}x^5][\mathcal{D}p_5]\exp\left(...\right)\rightarrow K_\mathrm{reg}(x-y; T)
\end{equation}
where $x-y$ and $0-nl_0$ are respectively  the four dimensional interval  and the separation along the fifth dimension. Already at this point, one can observe the regularity due to $l_0$, being
\begin{equation}
K_\mathrm{reg}(x-y; T)\sim\sum_n e^{(i\mu_0/2T)\left[(x-y)^2+n^2 l_0^2\right]}
\end{equation} 
where $\mu_0$ is a parameter which will not appear in the final result. 
Additional  integrations on $T$ and $w$ lead to Green's function
\begin{equation}
G_\mathrm{reg}(x-y)\sim \sum_w\int dT\ e^{(iT/2\mu_0)m_0^2} e^{(...w^2)} K_\mathrm{reg}(x-y; T),
\end{equation}
where $m_0$ is the mass of the particle in the limit $l_0\to 0$.
If one considers the leading term of the above expression, namely $n=w=1$, one finds  \eqref{eq:paddypropcoord} upon the condition 
$l_0=2\pi\sqrt{\alpha^\prime}$. In other words, the zero point length in four dimensions has a T-duality origin and coincides with the minimum length in string theory. 
The above result can be easily generalized to the case of  more than one compact dimension. The conclusion is unaffected: \eqref{eq:paddyprop} is both general and fundamental!

\subsection{How to implement T-duality effects}
\label{subsec:Tdualimpl}

Starting from \eqref{eq:paddypropcoord}, we expect important deviations from conventional Green's function equation 
\begin{equation}
\{\mathrm{Differential\ Operator}\}\ G(x,y)\ =\ \mathrm{Dirac\ Delta},
\label{eq:Greeneq}
\end{equation}
when $x\approx y $.  For the specific case of black holes, we recall that, in the absence of spin, there are both spherical symmetry and static conditions. It is, therefore, instructive to consider the 
interaction potential between two static sources with mass $m$ and $M$ due to \eqref{eq:paddyprop},
\begin{eqnarray}
V(r)
&=& -\frac{1}{m}\, \frac{W[J]}{T}\nonumber \\
&=& -GM\, \int\!\frac{d^3 k}{{\left(2\pi\right)}^3}\; {\left.G(k)\right|}_{k^0=0}\; 
 \exp\!\left(i \vec{k} \cdot \vec{r}\right)\nonumber \\
&=& -\frac{GM}{\sqrt{r^2 + l_0^2}}.
\label{eq:statpot}
\end{eqnarray}
The fact that $V\approx -GM/l_0$ for $r\to 0$, is the first signal of a possible removal of the curvature singularity. To verify  this is the case, one has to construct an effective energy momentum tensor for the r.h.s. of Einstein equations. The procedure is equivalent to the derivation of black hole solutions by means of non-local gravity actions
\begin{equation}
S=\frac{1}{2\kappa}\int \mathfrak{f}\left(R, \Box, \dots\right)\sqrt{-g}\ d^4 x + \int \mathfrak{L}\left( M , F^2, \Box, \dots \right)\sqrt{-g}\ d^4 x
\label{eq:fullaction}
\end{equation}
with $\kappa=8\pi G$, $\Box=\nabla_\mu \nabla^\mu$, $F$ is the gauge field and $\dots$ stand for higher derivative terms.

Eq. \eqref{eq:fullaction} is a compact notation for a class of actions that have been studied to obtain ghost free, ultraviolet finite gravity field equations \cite{Kra87,Tom97,Mod12,BGKM12}. For the present discussion, the 
details of such an action are not  relevant, since it is only an effective description of the full string dynamics. 
Accordingly, also the problem of the pathology of the action (e.g. ghosts, anomalies) is of secondary concerns, if one believes in the consistency of Superstring Theory. In conclusion, one can adopt a truncated version of the full non-local action \cite{Bar03,HaW05,Mof10,MMN11,GiN23} and derive the non-local Einstein equations.
For $F^2 =0$, they read
\begin{equation}
R_{\mu\nu}-\frac{1}{2}g_{\mu\nu} R= \kappa \ \mathfrak{T}_{\mu\nu}
\label{eq:nlee}
\end{equation}
where $\mathfrak{T}_{\mu\nu}={\cal O}^{-1}(\Box) T_{\mu\nu}$, while the Einstein tensor and  $T_{\mu\nu}$ are the conventional Einstein gravity tensors. The only thing that is important to know is the degree of ultraviolet convergence of the theory, encoded in the operator $\mathcal{O}(\Box)$. At this point, one can observe that \eqref{eq:statpot} is consistent with Green's function equation for \eqref{eq:paddypropcoord} \cite{GaN22}, namely
\begin{equation}
{\nabla ^2} G\left( {{\bf z},{{\bf z}^ \prime }} \right) =  - {l_0} \sqrt { - {\nabla ^2}}\, {K_1}\left( {{l_0}\sqrt { - 
{\nabla ^2}} } \right){\delta ^{\left( 3 \right)}}\left( {{\bf z} - {{\bf z}^ \prime }} \right).   \label{eq:staticgreen}
\end{equation}
The operator can be simply read off from the above equation, taking into account that $\mathcal{O}(\Box)=\mathcal{O}(\nabla ^2)$, if the source is static. In practice, the r.h.s. of \eqref{eq:staticgreen} is equivalent to the $\phantom{T}_t^t$ of the 
$\mathfrak{T}_{\mu\nu}$, namely
\begin{equation}
\mathfrak{T}_{t}^t\equiv -\rho(\mathbf{x})= (4\pi)^{-1} M {l_0} \sqrt { - {\nabla ^2}}\, {K_1}\left( {{l_0}\sqrt { - 
{\nabla ^2}} } \right){\delta ^{\left( 3 \right)}}\left( {\bf x}  \right).
\end{equation}
The effective energy density can be analytically derived and reads
\begin{equation}
\rho(\mathbf{x})=\frac{3l_0M}{4\pi\left(|\mathbf{x}|^2+l_0^2\right)^{5/2}}.
\end{equation}
For large distances, the above density quickly dies off as $\sim 1/|\mathbf{x}|^5$. Conversely, at short scales $|\mathbf{x}|\lesssim \l_0$, one finds the ``Sea of Tranquility'', i.e., a regular quantum region characterized by creation and annihilation of virtual particles at constant, finite energy. In such a sea, gravity becomes repulsive and prevents the full collapse of matter into a singularity.  With a geometric description in terms of differential line elements, the quantum fluctuations of such a sea are not visible. One can only capture the average effect, namely a local de Sitter ball around the origin, whose cosmological constant is $\sim GM/l_0^3$. Local energy condition violations certify the correctness of such a scenario.

After the above prelude, one can analytically solve \eqref{eq:nlee} and  display the full metric \cite{NSW19}
\begin{eqnarray}
ds^{2}=-\left(\,1-\frac{2M\LP^{2}\, r^2}{ (r^2
+l_0^2)^{3/2}} \,\right)dt^{2} +\left( \, 1-\frac{2M\LP^{2}\, r^2}{ (r^2
+l_0^2)^{3/2}} \, \right)^{-1}dr^{2}+r^{2}d\Omega^{2}.
\label{eq:Tdualitymetric}
\end{eqnarray}
The magic of the above result is that it coincides with the Bardeen solution \eqref{eq:Bardeenmetric}, provided
\begin{equation}
P\longrightarrow l_0.
\end{equation}
This is reminiscent of the relation between the holographic metric and the Reissner-Nordström geometry \eqref{eq:selfimplementation}:  this time, however, one can say that the Dirac string has been traded with a closed bosonic string. 

The general properties of the horizon structure and thermodynamics are similar to what seen in the context of the holographic metric -- see \ref{list:one}) -- \ref{list:four}) in Sec. \ref{subsec:holometric}. Horizon extremization allows for a SCRAM phase at the end of the evaporation, making the hole a stable system from a thermodynamic viewpoint. The Hawking temperature reads 
\begin{equation}
T= \frac{1}{4\pi\,r_{+}}\, \left(1 -\frac{3 \l_0^2}{r_{+}^2 +\l_0^2}\right),
\end{equation}
while the entropy is
\begin{eqnarray}
&& S
 = \frac{\mathcal{A}_{+}}{4} \left[ 
  \left(1 -\frac{8\pi\l_0^2}{\mathcal{A}_{+}}\right) \sqrt{1 +\frac{4\pi\l_0^2}{\mathcal{A}_{+}}}+\frac{12\pi\l_0^2}{\mathcal{A}_{+}} \left(\mathrm{arsinh}\sqrt{\frac{\mathcal{A}_{+}}{4\pi\l_0^2}} 
  -\mathrm{arsinh}\sqrt{2}\right)\vphantom{\sqrt{1 +\frac{\l_0^2}{r_{+}^2}}}\right],\nonumber \\ 
\end{eqnarray}
with  $\mathcal{A}_{+} = 4\pi r_{+}^2$. 

The great advantage of the metric \eqref{eq:Tdualitymetric} is the stability. This is a property in marked contrast to the case of the Bardeen metric, than can be, at the most a transient state. 
Even by postulating the existence of magnetic monopoles at some point of the history of the Universe \cite{Pre79}, their coupling has to be  much stronger than the QED coupling \cite{Pre84}
\begin{equation}
\alpha_\mathrm{m}\gg \alpha_e\sim 137^{-1}.
\end{equation}
This would imply for the Bardeen metric a sudden decay into the Schwarzschild black hole. 

Charged and charged rotating regular T-duality black holes have recently been derived. The novelty is the replacement of the ring singularity with a finite tension rotating string -- for further details see \cite{GJN22}.

\section{A short guide to black holes in noncommutative geometry}

We are going to present a family of black hole solutions, that represents a sort of coronation of the program of regular black holes in string theory. Indeed, after almost 20 years since their derivation, their good properties are still unmatched. 

One should start by recalling that noncommutative geometry (NCG) is a field in Mathematics whose goal is the  study of noncommutative algebras on certain topological spaces. In Physics, NCG has well known applications. For example, at the heart of quantum mechanics there is a noncommutative geometry, i.e.,  the algebra of quantum  operators. The idea that further physically meaningful results can be obtained from NCG, however, remained dormant at least until the 1990's. At that time,  Connes proposed the study of fundamental interactions from  the spectral triple principle \cite{Con95}.\footnote{The spectral triple is made of three items, a real, associative, noncommutative algebra $\mathcal{A}$, a Hilbert space $\mathcal{H}$ and a self adjoint operator $\eth$ on it.} 
As a  main goal, Connes aimed to construct a ``quantum version'' of the spacetime, by establishing a relation similar to that between  quantum mechanics and classical phase space \cite{Con96} (see also \cite{Sch05} for a pedagogical introduction).  

The most simple way to construct a noncommutative geometry is based on the
replacement of conventional coordinates with noncommutative operators
\begin{equation}
\left[x^i, \ x^j\right]=i\theta^{ij}
\label{eq:nccomm}
\end{equation}
where $\theta^{ij}$ is a constant, real valued, antisymmetric $D\times D$ matrix. The above commutator implies a new kind of uncertainty
\begin{equation}
\Delta x^i \Delta x^j \geq \frac{1}{2}\left|\theta^{ij}\right|,
\end{equation}
that can be used to improve the bad short distance behavior of fields propagating on the noncommutative geometry. To achieve this goal, one can deform field Lagrangians by introducing a suitable non-local product. For instance,  a  realization of noncommutive algebra of functions is based on the Moyal-produced (also known as star product or Weyl–Groenewold product)
\begin{equation}
f\star g \equiv \left. e^{(i/2)\theta^{ij}\frac{\partial}{\partial \xi^i}\frac{\partial}{\partial \eta^j}}f(x+\xi)g(x+\eta)\right|_{\xi=\eta=0},
\label{eq:starproduct}
\end{equation} 
that can be used as a starting point to obtain a noncommutative field theory  -- for reviews see e.g. \cite{Sza01,DoN01}.
Probably the biggest push to the popularity of noncommutative field theory was given by its connection to string theory.   Open strings ending on D-branes display a noncommutative behavior in the presence of a non vanishing, (constant) Kalb-Ramond $B$-field  \cite{SeW99}
\begin{equation}
\theta^{ij}\sim (2\pi \alpha^\prime)^2\left(\frac{1}{g+2\pi\alpha^\prime B}B\frac{1}{g-2\pi\alpha^\prime B}\right)^{ij},
\end{equation}
where $g$ is the metric tensor.

Noncommutative gravity follows a similar procedure for the metric field, defined over the underlying noncommutative manifold. The program of noncommutative gravity is, however, still in progress. Apart from some specific examples, one still misses a consistent noncommutative version of general relativity. In addition, the existing attempts to derive noncommutative corrections to classical black hole solutions run into the general difficulty of improving curvature singularities -- see \cite{Nic09}.

In 2003, the possibility of obtaining from noncommutative geometry something meaningful for the physics of black hole physics was still  perceived as quite remote. This was a time, that followed the ``explosive'' predictions about the possibility of a plentiful production of mini black holes in particle detectors \cite{DiL01,GiT02}. Operations at the LHC, however, began only five year later. As a result, there was a huge pressure to predict the experimental signatures of such black holes. If the terascale quantum gravity paradigm was correct, 
it was expected to have repercussions on  mini black hole cross section, evaporation and detection \cite{MNS12,NMSW15}.

Given this background, there was an unconventional attempt to study noncommutative geometry stripped of all elements, apart from its nonlocal character. From \eqref{eq:starproduct} one can guess that NCG introduces Gaussian damping terms.  To prove this, 
Smailagic and Spallucci considered the average of noncommutative  operators $\langle x^i \rangle$,
on states of minimal uncertainty, namely coherent states similar to those introduced by Glauber in quantum optics \cite{Gla63}. Such averages were interpreted as the closest thing to the conventional concept of coordinate. Initial results for path integrals on the noncommutative plane led to the conclusion that Dirac delta distributions are smeared out and become Gaussian functions, whose width is controlled by the noncommutative parameter $\theta$ \cite{SmSp03,SmSp03b}.\footnote{The matrix $\theta^{ij}$ in \eqref{eq:nccomm} can be written as $\theta^{ij}=\theta \varepsilon^{ij}$. The parameter $\theta$ has the dimension of a length squared.} The result was later formalized in terms of a nonlocal field theory formulation \cite{SSN06}. Green's function equation \eqref{eq:Greeneq} was determined by applying  a non-local operator to the source term namely\footnote{Here the signature is Euclidean.}
\begin{equation}
\delta^{(D)}(x-y)\longrightarrow f_\theta(x,y) = e^{\theta \Box} \delta^{(D)}(x-y).
\label{eq:ncsmearing}
\end{equation}

To derive a spacetime that account for noncommutative effects, one has to recall that the metric field can be seen as a ``thermometer'' that measures the average fluctuations of the manifold. 
From \eqref{eq:ncsmearing}, one can derive the effective energy density,
\begin{equation}
\rho(\mathbf{x})=\frac{M}{\left(4\pi\theta\right)^{3/2}} \ e^{-\left|\mathbf{x}\right|^2/4\theta},
\end{equation} 
and follow the procedure presented in Sec. \ref{subsec:Tdualimpl}. 
There are, however, two caveats: 
\begin{enumerate}
\setlength{\itemsep}{-\parsep}%
\item  
The resulting spacetime is  an \textit{effective description} that captures just one single character of NCG, i.e. non-locality.\footnote{For this reasons, one speaks of ``noncommutative geometry inspired'' solution. Other authors have termed it as ``minimalistic approach'' \cite{Vas09}.}    
\item The matrix $\theta^{ij}$ is assumed to behave like a field to preserve Lorentz symmetry.\footnote{Lorentz violation associated to \eqref{eq:nccomm} is a debated issue in the literature, e.g., see \cite{CHJK01,CCZ02,Mor03}. } 
\end{enumerate}

At this point, one can display the central result \cite{NSS06}
\begin{equation}
ds^{2}=-\left[\, 1-\frac{2M\LP^{2}}{r} \frac{\gamma\left(\frac{3}{2}; \frac{r^2}{4\theta} \right)}{\Gamma\left(\frac{3}{2}\right)} \,\right] dt^{2} +\left[\, 1-\frac{2M\LP^{2}}{r} \frac{\gamma\left(\frac{3}{2}; \frac{r^2}{4\theta} \right)}{\Gamma\left(\frac{3}{2}\right)} \,\right]^{-1}dr^{2}+r^{2}d\Omega^{2}.
\label{eq:NCSchw}
\end{equation}
Here $\gamma ( 3/2, x)\equiv \int_0^x\frac{du}{u} u^{3/2}e^{-u}$ is the incomplete Gamma function. 
It guarantees regularity of the manifold and quick convergence to the Schwarzschild metric for $r\gg\sqrt{\theta}$.
While the horizon structure and the thermodynamics are similar to those of the other quantum gravity improved metrics \eqref{eq:holometric}\eqref{eq:Tdualitymetric}, the above result has some specific characters.
 The Gaussian function \eqref{eq:ncsmearing} is a non-polynomial smearing, in agreement to what found by Tseytlin in \cite{Tse95}. On the other hand, polynomial functions (like GUP, T-duality) can be seen as the result of a  truncation of the expansion over the theta parameter \cite{KoN10}.  The above metric has been obtained also in the context of non-local gravity actions \cite{MMN11} and has been extended to the case of additional spatial dimensions \cite{Riz06,SSN09}, charged \cite{ANS++07} and rotating \cite{SmS10,MoN+10} solutions.  

From the emission spectra of the higher dimensional extension of \eqref{eq:NCSchw}, one learns that mini black holes tend to radiate soft particles mainly on the brane. This is in marked contrast with results coming from the Schwarzschild-Tangherlini metric \cite{NiW11}.

\section{Conclusions}

The very essence of the message I want to convey is the relation between particles and black holes in Fig. \ref{fig:fig1}.
It has already been noticed that strings and black holes share common properties \cite{tHo90}. In this work, however, the argument is reinforced and employed to improve classical black hole solutions. From this perspective the regularity of black hole metrics is the natural consequence of non-locality of particles, when described in terms of strings.

Another key point concerns the particle-black hole at the intersection of the curves for $\lambda$ and $r_\mathrm{g}$. 
The nature of this object is probably one of the most important topics in current research in quantum gravity. Indeed, the particle-black hole is essential to guarantee a self-complete character of gravity. Its mass and radius are related to the fundamental units of quantum gravity and string theory, along the common denominator of non-locality. This is evident also from the correspondence between cut offs, $\sqrt{\alpha^\prime}$ (string, GUP), $l_0$ (T-duality, GUP), $\LP$ self completeness, 
$\sqrt{\theta}$ (NCG).

In this work, we have also mentioned some of the existing difficulties, e.g., the details of the collapse at the Planck scale, the absence of an actual ``quantum manifold''. This means that the program of quantum gravity is far from being complete. 
It is also not clear if the predictions emerging from string theory will have experimental corroboration in the future. The ideas here presented, however, tend to support a less pessimistic scenario. Black holes could offer a testbed for fundamental physics, that is   alternative to conventional experiments in high energy particle physics.

\begin{footnotesize}

\subsubsection*{Acknowledgments}

The work of P.N. has partially been supported by GNFM, Italy's National Group
for Mathematical Physics. P.N. is grateful to Cosimo Bambi for the invitation to submit the present
contribution to  the volume ``Regular Black Holes: Towards a New Paradigm of Gravitational Collapse'', Springer, Singapore. P.N. is grateful to
Athanasios Tzikas for the support in drawing the picture.



\end{footnotesize}

\end{document}